\documentclass[letterpaper,notitlepage,nofootinbib,superscriptaddress]{revtex4-1} 

\usepackage[utf8]{inputenc}
\usepackage{amsmath,amssymb,amsfonts}
\usepackage{graphicx,wrapfig}
\usepackage{dcolumn}
\usepackage{bm}
\usepackage{color}
\usepackage{bbold}
\usepackage{url}
\usepackage{slashed}
\usepackage[svgnames]{xcolor}
\usepackage[english]{babel}
\usepackage{microtype}

\usepackage[colorlinks=true,linkcolor=blue,urlcolor=blue,citecolor=blue]{hyperref}

\usepackage{color}

\definecolor{darkgreen}{rgb}{0,0.65,0}

\newcommand{\be}{\begin{eqnarray}}
\newcommand{\ee}{\end{eqnarray}}
\newcommand{\ba}{\begin{array}}
\newcommand{\ea}{\end{array}}
\newcommand{\tr}{\mbox{\rm tr}}
\begin{document}
\title{\boldmath  Quadrupole pressure and shear forces inside baryons in the large $N_c$ limit }

\author{Julia  Yu.~Panteleeva}
	\affiliation{Ruhr University Bochum, Faculty of Physics and Astronomy,
Institute of Theoretical Physics II, D-44780 Bochum, Germany}
	\affiliation{Physics Department, Irkutsk State University, 
		Karl Marx str.~1, 664003, Irkutsk, Russia}
		\author{Maxim V.~Polyakov}
	\affiliation{Ruhr University Bochum, Faculty of Physics and Astronomy,
Institute of Theoretical Physics II, D-44780 Bochum, Germany}
\affiliation{Petersburg Nuclear Physics Institute, 
		Gatchina, 188300, St.~Petersburg, Russia}


\begin{abstract} 

We derive number of relations between quadrupole energy, elastic pressure, and shear force distributions in baryons using the large $N_c$ picture of  baryons as chiral solitons.
The obtained large $N_c$ relations are independent of particular dynamics and should hold in any picture in which the baryon is the chiral soliton. 

One of remarkable qualitative predictions of the soliton picture is the nullification of the tangential forces acting on the radial area element for any tensor polarisation of the baryon. 
The derived relations provide
a powerful tool to check the hypothesis that the baryons are chiral solitons,  say using lattice QCD.

\end{abstract}

\maketitle


\section*{Introduction}
\label{sec:Introduction}

The linear response of a hadron to a  change of the external space-time metric is described by the gravitational form factors (GFFs). For the first time the  GFFs for spin 0  and 1/2 were introduced and discussed in details
in Refs~ \cite{Kobzarev:1962wt,Pagels:1966zza}, for spin-1 particles in Ref.~~\cite{Holstein:2006ud} and for arbitrary spin hadrons in recent Ref.~\cite{Cotogno:2019vjb}.
The GFFs contain rich information about the internal structure of hadrons, for a detailed review see 
Ref.~\cite{Polyakov:2018zvc}. Particular interest for us here are the energy distributions and mechanical properties -- elastic pressure and shear force distributions
inside the hadron. These fundamental distributions are encoded in the static energy momentum tensor (EMT) defined in the Breit frame as \cite{Polyakov:2002yz}:
\be
\Theta^{\mu\nu} (\vec r, \sigma^\prime,\sigma ) 
= 
\int {d^3 \Delta \over  (2\pi)^3 2E } e^{-i \vec \Delta \cdot \vec r} 
\langle p^\prime, \sigma^\prime \, |{\hat \Theta}_{\rm QCD}^{\mu\nu}(0)|p,\sigma \rangle .
\ee
Here ${\hat \Theta}_{\rm QCD}^{\mu\nu}(0)$ is the QCD EMT operator which matrix element is computed between hadron states with spins projections $\sigma, \sigma'$ and 
momenta $p^0=p^{0\prime}=E=\sqrt{m^2+\vec \Delta^2/4}$,  and $p^{i\prime}=-p^i=\Delta^i/2$. The $00$ component of the static EMT contains the information
about the energy distribution inside the hadron, $0i$ components about the spin distribution, and $ik$ components provide us the distribution of elastic pressure and shear
forces inside the hadron  \cite{Polyakov:2002yz}. 

Various components of the static EMT for {\it arbitrary spin} hadron can be decomposed in multipoles of the hadron's spin operator. The expansion to the quadrupole
order has the following form \cite{Polyakov:2018rew,Polyakov:2019lbq,Sun:2020wfo}\footnote{In what follows, we shall suppress the hadron's spin indices $\sigma, \sigma'$ when their position is obvious.
Also we introduce here the parametrisation of the static stress tensor which differs from that in Ref.~\cite{Polyakov:2019lbq,Sun:2020wfo} by simple redefinition. 
The corresponding relations are given in Appendix. There we also collected some useful formulae.
}:
\be
\label{eq:quadrupole00}
\Theta^{00}({\bf r})& =&\varepsilon_0(r) +\varepsilon_2(r)\hat Q^{pq}Y_{2}^{pq}+\ldots,\\
\nonumber
\Theta^{ik}({\bf r})& =&p_0(r) \delta^{ik}+s_0(r)Y_2^{ik} + \left(p_2(r)+\frac13 p_3(r)-\frac19 s_3(r)\right) \hat{Q}^{ik} \\
\label{eq:quadrupoleik}
&+& \left( s_2(r)-\frac12 p_3(r)+\frac16 s_3(r) \right) 
2 \left[\hat{Q}^{ip}Y_{2}^{pk}+\hat Q^{kp}Y_{2}^{pi} -\delta^{ik} \hat Q^{pq}Y_{2}^{pq}    \right]\\
\nonumber
&+& \hat Q^{pq}Y_{2}^{pq}\left[\left(\frac23 p_3(r)+\frac19 s_3(r)\right)\delta^{ik}+\left(\frac12 p_3(r)+\frac56 s_3(r)\right) Y_2^{ik}\right]  +\ldots
\ee
Here ellipsis stays for the contribution of $2^n$- multipoles with $n>2$. The quadrupole operator is {the} $(2J+1)\times (2J+1)$
matrix:
\be
\hat Q^{ik}=\frac 12 \left( \hat J^{\ i} \hat J^{\ k}+\hat J^{\ k} \hat J^{\ i} -\frac 23 J(J+1) \delta^{ik} \right),
\ee
which is expressed in terms of the spin operator $\hat J^{\ i}$. The spin operator can be expressed in terms of the SU(2) Clebsch-Gordan coefficients (in the spherical basis):
\be
\hat J^{\ \mu}_{\sigma'\sigma}= \sqrt{J(J+1)}\  C_{J \sigma 1\mu}^{J \sigma'}.
\ee
Also we introduce the irreducible (symmetric and traceless) tensor  of $n$-th rank:
\be
Y_{n}^{i_1 i_2 ... i_n} = \frac{(-1)^n}{(2 n-1)!!} r^{n+1} \partial^{i_1}...\partial^{i_n} \frac{1}{r}, 
\quad \mbox{i.e.} \quad
Y_0=1,\ Y_1^{i}=\frac{r^{i}}{r}, \ Y_2^{ik}=\frac{r^{i} r^{k}}{r^2}-\frac13 \delta^{ik}, \ {\rm etc.}
\ee 
Note that only monopole quantities $\varepsilon_0(r)$, $p_0(r)$, and $s_0(r)$ are left after the spin average.
The functions $\varepsilon_0(r)$ and $\varepsilon_2(r)$ correspond to the spin averaged energy density and  to the quadrupole deformation of the energy density in the hadron
correspondingly. There is obvious relation $\int d^3 r\ \varepsilon_0(r)=m $. Also it is obvious that $\varepsilon_2(r)=0$ for the hadrons of spin 0 and 1/2. That is why such hadrons can be
called spherically symmetric.

From the stability condition for the stress tensor $\partial_i \Theta^{ik}({\bf r})=0$ one can easily obtain the equations for the functions $p_n(r)$ and $s_n(r)$:

\be
\label{eq:SFE}
\frac{d}{dr} \left(p_n(r)+\frac 23 s_n(r)\right)+\frac 2r s_n(r)=0, \ \ {\rm for }\ n=0,2,3.
\ee 
These equations\footnote{
Such type of the equation can be called  ``hadron shape formation equation".
Indeed, the non-trivial shape of the pressure distribution (hadron shape) 
appear due to non-trivial shear force distribution $s_n(r)$, the latter is 
also called  pressure anisotropy \cite{Lorce:2018egm}. Interestingly 
the pressure anisotropy (shear force distribution) plays an essential 
role in astrophysics \cite{Lorce:2018egm}, see the review 
\cite{Herrera:1997plx} on the role of pressure {anisotropy}
for self-gravitating systems in astrophysics and cosmology. 
} 
 have the form of the equilibrium relation between the elastic pressure distribution $p_0(r)$ and the shear force distribution  $s_0(r)$ for
 spherically symmetric systems, see
e.g. Refs~\cite{Polyakov:2002yz,Polyakov:2018zvc}. Therefore, we call the functions $p_2(r)$, $p_3(r)$ as the quadrupole elastic pressure distributions, and
the functions $s_2(r)$, $s_3(r)$ as the quadrupole shear force distributions. The functions  $p_0(r)$, $s_0(r)$ correspond to the spin averaged pressure and shear force
distributions, they coincide with the distributions for spherically symmetric hadrons of the spin 0 and 1/2.

The solution of the Eq.~(\ref{eq:SFE}) can be written in terms of 
the 3D Fourier transform of (generalised) D-form factors:
\be
\label{eq:quadrupoleps}
p_n(r)=\frac{1}{6m} \frac{1}{r^2} \frac{d}{dr}r^2\frac{d}{dr} \widetilde{D}_n(r)=\frac{1}{6m} \partial^2\   \widetilde{D}_n(r),\quad s_n(r)=-\frac{1}{4m} r \frac{d}{dr}\frac{1}{r}\frac{d}{dr} \widetilde{D}_n(r).
\ee
The form (\ref{eq:quadrupoleps}) of the quadrupole pressure and shear forces also ensures that all relations 
for the force distributions discussed in Sec.~IX and App.
of Ref.~\cite{Polyakov:2018zvc} are satisfied automatically. In particular, the (generalised) von Laue conditions are satisfied automatically:
\be
\int d^3 {\bf r} \, p_n(r) =\frac{1}{6 m}\int d^3 {\bf r}\ \partial^2\   \widetilde{D}_n (r)=0 \ , {\rm with }\  n=0,2,3 \  .
\ee
 Note that the dimensionless constants  (generalised D-terms):
  \be
  \label{eq:dterms}
 {\mathcal D}_n\equiv \int d^3 {\bf r}\  \widetilde{D}_n(r)=m  \int d^3 {\bf r}\  r^2\ p_n(r)={- \frac{4}{15}\,m \int d^3 {\bf r}\;r^2s_n(r)} \,  ,
 \ee 
 are  characteristics of the elastic properties of the hadron which are as fundamental as other mechanical properties
 of the hadron such as the mass and the spin. In principle, they could be listed in PDG on equal footing with the mass and spin of particles. The first measurements
 of ${\mathcal D}_0$
in hard QCD processes became available  for the nucleon in Refs.~\cite{Kumericki:2015lhb,Nature} and in Ref.~\cite{Kumano:2017lhr} for the pion. Profound studies of all
subtleties in extraction of the D-term ${\mathcal D}_0$ from hard exclusive processes can be found in Ref.~\cite{Kumericki:2019ddg}.
 
 The first studies of the quadrupole energy, elastic pressure, and shear force distributions were performed  in Ref.~\cite{Sun:2020wfo}  for the case of $\rho$-meson, where, the authors employed the light-cone constituent quark model.  In the present paper we shall derive the relations between  quadrupole energy, elastic pressure, and shear force distributions
 for the baryons in the large-$N_c$ limit. In the latter limit the baryons can be viewed as the chiral solitons. Our relations are independent of the dynamics (effective field theory)
 describing the chiral soliton and can be used as a strong criterion to check the hypothesis that the baryons are chiral solitons.

\section*{\boldmath Gravitational form factors of the baryon as chiral soliton}
\label{Sec-5:Skyrme}

The most striking success of the old Skyrme idea \cite{Skyrme:1961vq} that baryons can be viewed as solitons of the pion (or chiral) field, is the classification of light baryons it suggests.
This idea implies that various baryons are quantum excitations of the same classical object -- the chiral soliton and, hence, the properties of baryons are interrelated.
Quantum Chromodynamics has shed some light into why the chiral soliton picture is correct: we know now that the spontaneous chiral symmetry breaking in QCD is, probably, the most important feature of strong interactions, determining to a great extent their dynamics, while the large $N_c $(= numbers of colours) argumentation by Witten \cite{Witten:1979kh,Witten:1983tx} explains why the pion field inside the nucleon can be considered as a classical one, i.e. as a ``chiral soliton".

Following Witten \cite{Witten:1983tx}  we assume the self-consistent pseudoscalar field which binds up the $N_c$ quarks in the ``classical" baryon (i.e. the soliton field) to be of the 
hedgehog form\footnote{We consider the case of two flavours and the small violation of the isospin symmetry is neglected. The generalisation to the three flavour case and
inclusion of the flavour symmetry breaking terms are straightforward.}:

\be
\label{eq:hedgehog}
U_0({\bf  r})=\exp \left( i \tau^a   n^a P(r)\right),
\ee
where the unit vector $n^a=r^a/r $, and  the spherically-symmetric profile function $P(r)$ is defined by dynamics. We shall not need the concrete form of this function in what follows--
for us the particular form of the underlying effective field theory is not relevant. The only hypothesis we do here is that the baryon is the chiral soliton of the form (\ref{eq:hedgehog}).

In order to provide the ``classical" baryon with specific quantum numbers one has to
consider an $SU(2)$-rotated pseudoscalar field:
\be
U({\bf  r},t)=R(t) U_0({\bf  r}) R^\dagger(t),
\ee
where $R(t)$ is an unitary $SU(2)$ matrix depending only on time and $U_0({\bf  r})$ is the static hedgehog field given by Eq.~(\ref{eq:hedgehog}). 
Due to chiral symmetry the dynamics can depend only on the angular velocity of the rotation: 
\be
\Omega^i= \frac i2 \tr\left ( R\partial_t R^\dagger \tau^i\right).
\ee
Quantizing this rotation one gets the spectrum of baryons and relation of the angular velocity to the spin operator of the corresponding baryon $\Omega^i =\hat{J^i}/I$.
Here $I\sim N_c$ is the soliton moment of inertia, its particular value is not relevant for us here. We see that the expansion of any observable in the angular velocity corresponds to the $1/N_c$ expansion. 

Looking on Eqs.~(\ref{eq:quadrupole00},\ref{eq:quadrupoleik}) we come to our first conclusion that all quadrupole quantities appear in the second order of the angular velocity expansion and, hence,
are $1/N_c^2$ suppressed relative to monopole one. Moreover,  we make a key observation -- the angular velocity dependence can enter any quantity only through the zero components of the left and right 
chiral currents:

\be
L_0&=&U^\dagger ({\bf  r},t) \partial_t U({\bf  r},t)=U^\dagger({\bf  r},t) \left[U({\bf  r},t),\Omega \right]\propto  i U^\dagger({\bf  r},t) \left[\vec \Omega  \times {\vec n}\right]^i \tau^i,\\
\nonumber
R_0&=&U({\bf  r},t) \partial_t U^\dagger ({\bf  r},t)=U({\bf  r},t) \left[U^\dagger ({\bf  r},t),\Omega \right]\propto  i U({\bf  r},t) \left[\vec \Omega  \times {\vec n}\right]^i \tau^i,\\
\ee
where the last proportionality follows from the hedgehog form of the chiral field of the soliton (\ref{eq:hedgehog}). It reflects the fact that for the hedgehog form of
the chiral field the isospin rotations can be compensated by the rotation of the coordinate system. 
From this key observation we conclude that
the static EMT in the soliton picture  can depend on the baryon's spin operator $\hat{J}^i$ only through the vector product
$\left[ \hat{\vec J}  \times {\vec n}\right]^i $, independently of concrete dynamics.

\subsection*{Energy densities of rotating chiral soliton}
\label{Sec-5a:Skyrme-with-Nc-corrections}

Above we observed that the $1/N_c$ corrections to the static EMT can be  obtained as the expansion in the vector $\left[\vec \Omega  \times {\vec n}\right]^i $.
Therefore we can write general form of the rotational corrections (up to $\sim \Omega^2$ order) to the static $\Theta^{00}({\bf r})$ as:
\be
\label{eq:t00rot}
\delta_{\rm rot} \Theta^{00}({\bf r})=\left[\vec \Omega  \times {\vec n}\right]^i \left[\vec \Omega  \times {\vec n}\right]^i\ F(r).
\ee
Here the function of the radial coordinate $F(r)$ depends on the concrete dynamics and again not relevant for derivations here.
Quantizing the rotations ($\Omega^i=\hat{J^i}/I$) and comparing the obtained form to the general  parametrisation  (\ref{eq:quadrupole00}) we obtain the first 
relation:

\be
\label{r0}
\delta_{\rm rot} \varepsilon_0^{(J)}(r)=-\frac 23 J(J+1)\ \varepsilon_2(r).
\ee
Here the rotational correction to the monopole energy density $\delta_{\rm rot} \varepsilon_0^{(J)}(r)$ for the baryon excitation of the spin $J$ has general form
$\delta_{\rm rot} \varepsilon_0^{(J)}(r)\sim J(J+1)$ which is shown in Eq.~(\ref{r0}).
Note
that this relation (and the relations derived below) can have  corrections of the order of  $\sim 1/N_c^3$.
Using the relation (\ref{r0}), we can relate the energy densities for $\Delta$ baryon ($J=3/2$) and the nucleon in the following way:

\be
\label{r1}
\varepsilon^\Delta_0(r)+2 \varepsilon^\Delta_2(r)=\varepsilon^N_0(r).
\ee
This is the first example of relations between mechanical characteristics of different baryons which follows from the soliton nature of baryons in the large $N_c$ limit. 
To estimate the typical size of the quadrupole energy density $\varepsilon^\Delta_2(r)$ we first use the obvious relation:
\be
\int d^3 {\bf r}\ \left( \varepsilon^\Delta_0(r)-\varepsilon^N_0(r)\right)=m_\Delta-m_N.
\ee
Further, with help of Eq.~(\ref{r1}), we obtain:
\be
\int d^3 {\bf r}\  \varepsilon^\Delta_2(r)=-\frac 12 (m_\Delta-m_N)\approx -146\ {\rm MeV}.
\ee
This numerical value is about 10\%  of  the integral $\int d^3 {\bf r}\  \varepsilon^\Delta_0(r)=m_\Delta\approx 1232~{\rm MeV}$, compatible with being the $1/N_c^2$ correction.

\subsection*{\boldmath  Quadrupole pressure and shear forces of rotating chiral soliton}

Again using the fact that the dependence of the static EMT of the chiral soliton can depend on the angular velocity only through the vector   $\left[\vec \Omega  \times {\vec n}\right]^i $,
we can write the general decomposition of the angular velocity correction to static stress tensor as:

\be
\label{eq:tikrot}
\delta_{\rm rot} \Theta^{ik}({\bf r})=\left[\vec \Omega  \times {\vec n}\right]^i \left[\vec \Omega  \times {\vec n}\right]^k\ G_1(r)+
\left[\vec \Omega  \times {\vec n}\right]^p \left[\vec \Omega  \times {\vec n}\right]^p \left[ \delta^{ik} G_2(r)+Y_2^{ik} G_3(r)\right].
\ee
Here $G_{1,2,3}(r)$ are functions depending on the concrete dynamics. Quantizing the rotations ($\Omega^i=\hat{J^i}/I$) and comparing the obtained form  to the 
general parametrisation  (\ref{eq:quadrupoleik}) we obtain remarkable relation:

\be
\label{eq:first}
p_2(r)+\frac 23 s_2(r)=0.
\ee 
The combination $p_2(r)+\frac 23 s_2(r)$ enters the equilibrium equation (\ref{eq:SFE}) which under the condition (\ref{eq:first}) implies nullification
of both $p_2(r)$ and $s_2(r)$. Taking into account this nullification we arrive eventually to the following non-trivial relations:

\be
\label{r2}
p_2(r)=s_2(r)=0, \ \delta_{\rm rot} s_0^{(J)}(r)=-\frac 23 J(J+1)\ s_3(r) ,\ \delta_{\rm rot} p_0^{(J)}(r)=-\frac 23 J(J+1)\ p_3(r) .
\ee
The first of these relations is very interesting predictions of the large $N_c$ picture of baryons as the chiral solitons. Due to its simplicity it is the easiest to check, say on the lattice or in other QCD based models.
These would be the nice check of the soliton nature of baryons. 

Let us see the physics meaning of $p_2(r)$ and $s_2(r)$ which nullify in the soliton picture of baryons.
With the parameterisation (\ref{eq:quadrupoleik}) the force acting on the infinitesimal 
radial area element $dS_r$  ($d\bm{S}=dS_r\bm{e}_r+dS_\theta\bm{e}_\theta+dS_\phi\bm{e}_\phi$)
has the following spherical components:
\be\label{Eq:force-spherical components}
	\frac{dF_r}{dS_r}&=&p_0(r) + \frac 23 s_0(r)+\hat Q^{r r}  \left(p_2(r)+ \frac 23 s_2(r)+p_3(r)+\frac 23 s_3(r)\right) , \\	
	\label{Eq:force-spherical components1} 
	\frac{dF_\theta}{dS_r}&=& \hat Q^{\theta r}  \left(p_2(r)+ \frac 23 s_2(r)\right), \quad \frac{dF_\phi}{dS_r}= \hat Q^{\phi r}  \left(p_2(r)+ \frac 23 s_2(r)\right).
\ee
We see that in contrast to the spherically symmetric hadron, the radial area element experiences not only normal forces, but the tangential one as well. 
The size of the tangential forces are governed by $p_2$ and $s_2$ as these forces
are proportional to $ \frac 23 s_2(r)+p_2(r)$. The absence of these tangential forces is the remarkable prediction of 
the soliton picture of baryons.

With help of  Eq.~(\ref{r2}) we can relate the fundamental
characteristics of the baryon elastic properties  (\ref{eq:dterms}) (generalised D-terms) for different baryons:

\be
\label{r3}
{\mathcal D}^\Delta_2=0, \ {\mathcal D}^\Delta_0+2 {\mathcal D}^\Delta_3={\mathcal D}^N_0.
\ee
The same relations are valid for pressure and shear force distributions  in the nucleon and in $\Delta$ baryon.

The rotational corrections to the monopole (spin averaged) pressure and shear force distributions were computed in the framework of the Skyrme model
in Ref.~\cite{Cebulla:2007ei} and studied in the same model in details in Ref.~\cite{Perevalova:2016dln}. We refer the reader to these papers for the numerical values
of the rotational corrections in the Skyrme model. Here we just give the value of the generalised D-terms (\ref{eq:dterms}) for $\Delta$ baryon and the nucleon which we 
obtain using the Skyrme model
 results of Ref.~\cite{Perevalova:2016dln}:

\be
{\mathcal D}^N_0\approx -3.40,\ {\mathcal D}^\Delta_0\approx -2.65,\  {\mathcal D}^\Delta_2=0,\ {\mathcal D}^\Delta_3\approx -0.38. 
\ee
The relative numerical values are compatible with the expectation from the $1/N_c$ counting.

\section*{Conclusion and outlook}
\label{Sec-7:conclusions-outlook}

We derive number of relations between quadrupole energy, elastic pressure, and shear force distributions in baryons using the large $N_c$ picture of  baryons as chiral solitons
(see Eqs.~(\ref{r0},\ref{r1},\ref{r2},\ref{r3})).
The obtained large $N_c$ relations are independent of particular dynamics and should hold in any picture in which the baryon is the chiral soliton. The relations provide
a powerful tool to check the hypothesis that the baryons are chiral solitons,  say using lattice QCD.

Probably the most remarkable (and the easiest to check) prediction of the soliton picture is the nullification of the generalised D-term ${\mathcal D}_2$. We think it 
should be rather easy to measure it on the lattice.
Qualitatively this  prediction of the soliton picture implies the nullification of the tangential forces acting on the radial area element for any tensor polarisation of the baryon,
see Eqs.~(\ref{Eq:force-spherical components1},\ref{Eq:force-spherical components_dr1}). 
It might be that this nullification is
 more general requirement, like the nullification of the anomalous gravitomagnetic moment for spin 1/2 fermions proven in \cite{Kobzarev:1962wt}. The proof of the nullification of the
 the anomalous gravitomagnetic moment for particles of any spin is given in Ref~\cite{Teryaev:2016edw}.
 At the moment we are not able to prove the conjecture that $p_2(r)$ and $s_2(r)$ are zero  beyond $1/N_c$ expansion.  

In the consideration here we restrict ourselves to two-flavour QCD and to the first rotational band of baryons like the nucleon and $\Delta$ baryon. We think it is pretty 
straightforward to make the generalisation to the three-flavour case (see, e.g.  Ref.~\cite{Pant}). Also the generalisation to the case of other rotational bands (excited resonance)
can be performed with help of methods developed in Ref.~\cite{Diakonov:2013qta}.


\acknowledgments

JP and MVP are thankful to S.~E.~Korenblit and I.~A.~Perevalova for conversations which initiated presented studies. 
M.V.P.\ is grateful to Peter Schweitzer and Bao-Dong Sun for 
many illuminating discussion.
This work was supported in part the Deutsche Forschungsgemeinschaft
(CRC110) and by the BMBF (grant 05P2018).

\appendix

\section*{Appendix }

We give here the conversion of quadrupole elastic pressure and shear forces used in Refs.~ \cite{Polyakov:2019lbq,Sun:2020wfo}
to those used here, see Eq.~(\ref{eq:quadrupoleik}):

\be
p_0(r)&=&\bar{p}_0(r),\\ 
s_0(r)&=&\bar{s}_0(r), \\
p_2(r)&=&\bar{p}_2(r)-\dfrac{1}{m^2}\left(-\dfrac{2}{3}\bar{p} ''_3(r)+\dfrac{2}{9}\bar{s}''_3(r)+\dfrac{2}{3}\dfrac{\bar{p}'_3(r)}{r}-\dfrac{6}{27}\dfrac{\bar{s}'_3(r)}{r}+2\dfrac{\bar{s}_3(r)}{r^2}\right), \\
s_2(r)&=&\bar{s}_2(r)+\dfrac{1}{m^2}\left(\dfrac{1}{3}\bar{s} ''_3(r)-\bar{p}''_3(r)+\dfrac{\bar{p}'_3(r)}{r}-\dfrac{10}{3}\dfrac{\bar{s}'_3(r)}{r}+6\dfrac{\bar{s}_3(r)}{r^2}\right), \\
p_3(r)&=&\dfrac{1}{m^2}\left(\dfrac{2}{9}\bar{s} ''_3(r)-\dfrac{5}{3}\bar{p}''_3(r)+\dfrac{5}{3}\dfrac{\bar{p}'_3(r)}{r}-\dfrac{20}{9}\dfrac{\bar{s}'_3(r)}{r}+4\dfrac{\bar{s}_3(r)}{r^2}\right),\\ 
s_3(r)&=&\dfrac{1}{m^2}\left(\bar{p} ''_3(r)-\dfrac{4}{3}\bar{s}''_3(r)-\dfrac{\bar{p}'_3(r)}{r}+\dfrac{22}{3}\dfrac{\bar{s}'_3(r)}{r}-12\dfrac{\bar{s}_3(r)}{r^2}\right).
\ee
Here the functions $\bar{p}_n(r)$ and $\bar{s}_n(r)$ correspond to quadrupole pressure and share force distributions from Refs.~\cite{Polyakov:2019lbq,Sun:2020wfo},
and  ${p}_n(r)$ and ${s}_n(r)$ corresponding quantities from Eq.~(\ref{eq:quadrupoleik}). We remind that $\bar{p}_n(r)$ and $\bar{s}_n(r)$  from Refs.~\cite{Polyakov:2019lbq,Sun:2020wfo}
also satisfy `` hadron shape forming" (equilibrium) equation (\ref{eq:SFE}), one can easily check that the conversion formulae given above preserve the form of the equilibrium equation.

The physics meaning of the quadrupole elastic pressures ($p_2$ and $p_3$) and quadrupole shear forces ($s_2$ and $s_3$) is illustrated by 
Eqs.~(\ref{Eq:force-spherical components},\ref{Eq:force-spherical components1}).
Yet another way to show that  the equilibrium equations (\ref{eq:SFE}) can be viewed as ``hadron shape formation" equation is given by simple fact that
the change of the forces with the distance from the hadron centre is governed by:
\be
\label{Eq:force-spherical components_dr}
\frac{d}{dr}\left(	\frac{dF_r}{dS_r}\right)&=&-\frac{2}{r} \left[s_0(r)) + \hat Q^{r r}  \left(s_2(r)+ s_3(r)\right) \right] , \\
\label{Eq:force-spherical components_dr1}	 
\frac{d}{dr}\left(	\frac{dF_\theta}{dS_r}\right)&=& -\frac{2 s_2(r)}{r}\hat Q^{\theta r}, \quad  \frac{d}{dr}\left(\frac{dF_\phi}{dS_r}\right)= -\frac{2 s_2(r)}{r} \hat Q^{\phi r}.
\ee

\end{document}